\documentclass[fleqn,usenatbib]{mnras}

\usepackage[T1]{fontenc}
\usepackage{hyperref}
\usepackage{xcolor}

\DeclareRobustCommand{\VAN}[3]{#2}
\let\VANthebibliography\thebibliography
\def\thebibliography{\DeclareRobustCommand{\VAN}[3]{##3}\VANthebibliography}

\usepackage{graphicx}
\usepackage{amsmath}	
\usepackage{amssymb}	
\usepackage{xspace}
\usepackage{subfigure}

\newcommand{\msolar} {$\rm{M_{\odot}}~$}
\newcommand{\msolarc} {$\rm{M_{\odot}}$}
\newcommand{\kms} {km $\rm{s^{-1}}$}

\usepackage{newtxtext,newtxmath}

\title[MBH progenitor evidence in Leo I]{Observational Signatures of Massive Black Hole Progenitor Pathways: Could Leo I be a Smoking Gun?}

\author[J.A. Regan, F. Pacucci \& M. J. Bustamante-Rosell]{John A. Regan$^{1}$\thanks{E-mail:john.regan@mu.ie, Royal Society - SFI University Research Fellow},
Fabio Pacucci$^{2,3}$ \& M. J. Bustamante-Rosell$^4$
\\
$^1$Centre for Astrophysics and Space Science Maynooth, Department of Theoretical Physics, Maynooth University, Maynooth, Ireland\\
$^2$Center for Astrophysics, Harvard \& Smithsonian, Cambridge,  MA 02138 USA\\
$^3$Black hole Initiative,  Harvard University, Cambridge, MA 02138 USA\\
$^4$Department of Astronomy and Astrophysics, University of California, Santa Cruz, CA 95064, USA}

\date{\today}

\pubyear{2022}

\begin{document}
\label{firstpage}
\pagerange{\pageref{firstpage}--\pageref{lastpage}}
\maketitle

\begin{abstract}
\noindent Observational evidence is mounting regarding the population demographics of Massive Black Holes (MBHs), from the most massive cluster galaxies down to the dwarf galaxy regime. However, the progenitor pathways from which these central MBHs formed remain unclear. Here we report a potentially powerful observational signature of MBH formation in dwarf galaxies. We argue that a continuum in the mass spectrum of MBHs in (fossil) dwarf galaxies would be a unique signature of a heavy seed formation pathway. The continuum in this case would consist of the usual population of stellar mass black holes, formed through stellar evolution, plus a smaller population of heavy seed MBHs which have not yet sunk to the centre of the galaxy. Under the robust assumption of initial fragmentation of the parent gas cloud resulting in a burst of heavy seed production, a significant fraction of these seeds will survive to the present day as off-nuclear MBHs with masses less than that of the central object. Motivated by the recent discovery of a MBH in the relatively low central density Leo I galaxy, we show that such a continuum in MBH seed masses should persist from the lightest black hole masses up to the mass of the central MBH in contrast to the light seeding scenario where no such continuum should exist. The detection of off-centered MBHs and a central MBH would represent strong evidence of a heavy seeding pathway.
\end{abstract}

\vspace{-0.5cm}
\section{Introduction} \label{Sec:Introduction}
\noindent Dwarf galaxies are typically defined as having stellar masses below $3 \times 10^{9}$ \msolarc. In a cosmological context, they have become increasingly important in recent years as they resemble the earliest galaxies formed at high redshift, and some may be the fossil remnants of these very early 
galaxies \citep[e.g.][]{Bovill_2011, Frebel_2014, Collins_2022}. Additionally, whether or not these small galaxies host central massive black holes (MBHs) has
been a topic of focused investigation over the last decade or so. Initial research into using (fossil) dwarf galaxies to understand the formation mechanisms of MBHs at high redshift was pioneered by \cite{Volonteri_2008} and \cite{VanWassenhove_2010} with a significant observational focus now taking place on determining the occupation fraction of MBHs in dwarf galaxies in the present day Universe \citep[e.g.][]{Baldassare_2020}.\\
\indent Detecting and determining the occupation fraction of MBHs in dwarf galaxies remains a significant challenge, with the 
occupation fraction and the active fraction currently unknown and debated \citep{Pacucci_2021}. 
Most searches of dwarf galaxies thus far have focused on using optical narrow emission line diagnostic diagrams to identify active galactic nuclei (AGN) emission and broad emission lines to estimate the MBH mass \citep{Greene_2004, Greene_2007, Reines_2013, Moran_2014, Chilingarian_2018}. Additional searches in the X-ray have also revealed numerous candidate AGN in dwarf galaxies out to much higher redshift \citep[][]{Pardo_2016, Mezcua_2018, Mezcua_2019, Mezcua_2020}
However, these and similar techniques are subject to high systematic uncertainties, and a cleaner method for determining the existence and mass of MBHs in dwarf galaxies comes from the kinematics of stars. 
Unmediated by gas dynamics, stellar velocity measurements can give an unbiased probe of the gravitational potential in the central parsecs of the host galaxy. 
Resolving the gravitational effect of a MBH requires kinematic measurements within its sphere of influence \citep{Peebles_1972}, which has been limited to relatively nearby galaxies. 
The pioneering work of \cite{Kormendy_1995} has been extended to additional, nearby galaxies by, for example, \cite{McConnell_2012} and \cite{Liepold_2020}.\\
\indent The kinematic method does not measure the MBH mass directly but rather the total gravitational potential of the host and any MBH within the host galaxy \citep{DenBrok_2014, Thater_2017, Nguyen_2018, Nguyen_2019}. 
Hence, kinematics at several radii, a luminosity profile, and dynamical modelling are necessary to separate the mass components of the galaxy \citep[e.g.][]{1998ApJ...493..613V,2000AJ....119.1157G,2002ApJ...578..787C,2003ApJ...583...92G,2013AJ....146...45R}.
One of the few systematic uncertainties of the method is in the dynamical modelling procedure — the most computationally expedient methods (e.g. Jeans analysis) assume a known form for the velocity anisotropy and dark matter profile. In principle, these restrictions are avoidable with non-parametric modelling, albeit at a much higher computational cost. Using this methodology \cite{Bustamante-Rosell_2021} recently determined that
the Leo I dwarf galaxy contains a MBH with a mass of M$_{MBH} = (3.3 \pm 2) \times 10^6$ \msolarc. \\
\indent In this letter, we use the Leo I result 
together with analytic arguments and findings from high-z simulations to argue that dwarf spheroidal galaxies, as well as similar low-density dwarf galaxies, potentially host a previously unexplored signature of MBH seeding pathways. \\
\indent \textit{Light} seeds (those emerging from the remnants of the very first stars \citep{Madau_2001}) could be the progenitors for MBHs — in order to do so, they would have to grow extremely efficiently - something that so far appears challenging to achieve in practice \citep[e.g.][]{Smith_2018}. \textit{Heavy} seeds, on the other hand, are thought to be born with masses, possibly via an intermediate stage as a super-massive star \citep{Woods_2017}, in the range M$_\mathrm{seed} \simeq 10^{3 - 5}$ \msolar in high-z galaxies that resemble today's dwarfs. \\
\indent For the purposes of this paper we use the term \textit{heavy} seed for all masses greater than $10^3$ \msolarc. We are cognisant this is in tension with some nomenclature which would 
instead refer to black holes with masses of approximately $10^3$ \msolar as "medium" weight seeds and only those greater than approximately $10^4$ as \textit{heavy} seeds. The mass of the medium weight seeds is a robust prediction of dynamical models of MBH formation  \citep[e.g.][]{Miller_2012, Katz_2015, Stone_2017, Schleicher_2022} which either through runaway stellar collisions or through the repeated mergers of lighter black holes produce black holes with masses of approximately $10^3$ \msolarc. However, more recently this distinction (in resulting black hole masses) is becoming blurred with both simulations by \cite{Chon_2020} and \cite{Regan_2020b} predicting initial black hole masses in the range $10^{3-4}$ \msolar due to certain environmental dependencies, which previously were thought to produce \textit{heavy} seeds. Perhaps the more fundamental difference between the scenarios is that in the 
model scenarios of \cite{Chon_2020} and \cite{Regan_2020b} a significant number (and spectrum) of black hole masses is predicted due to fragmentation of the parent gas cloud. In contrast the dynamical pathways predict a single MBH with a mass in the range $10^3$ \msolarc. Therefore, the signature we postulate here should be a unique signature of a scenario in which multiple \textit{heavy} seeds are formed from fragmentation.\\
\indent In summary our proposition here is that \textit{heavy} seeds born at high redshift, through either rapid halo assembly or similar processes, are typically formed in multiples, due to modest fragmentation of the parent gas cloud. In fossil dwarf galaxies that don't have overly dense central structures (i.e., they are below the density typical of nuclear star clusters (NSCs)), a significant number of these initial fragments will survive and constitute a robust observational signature of the initial seeding pathway.\\
\indent In \S \ref{Sec:LeoI} we discuss the characteristics of the Leo I galaxy and its MBH.
In \S $\ref{Sec:Model}$ we outline models for MBH growth through both the \textit{light} and \textit{heavy} seed channels, showing how the different pathways may be distinguished given sufficiently sensitive observations of MBH demographics in fossil dwarf galaxies. 
In \S \ref{Sec:Discussion} we discuss the broader implications of our postulates and give our conclusions. 

\section{The Massive Black Hole in the Dwarf Galaxy \textit{Leo I}} \label{Sec:LeoI}
\noindent
The recent detection of a MBH at the centre of the dwarf spheroidal galaxy Leo I by \cite{Bustamante-Rosell_2021} represents one of the most remarkable MBH discoveries to date. Its mass was estimated at M$_{MBH} = (3.3 \pm 2) \times 10^6$ \msolar lifting
it significantly above the standard M$_{MBH} - \sigma$ relation \citep{Kormendy_2013, Baldassare_2020, Greene_2020} for both very massive and dwarf galaxies alike. 

Prior studies of Leo I used individual stellar kinematics and stellar counts to probe the gravitational potential of the dwarf spheroidal \citep{Mateo_2008, Sohn:2006et, Koch:2007ye}.
\cite{Bustamante-Rosell_2021} showed that when concentrated in the central parsecs of the galaxy, individual stellar kinematics suffered from crowding, which biased this method towards inferring lower velocity dispersions, which in turn led to inferring lower enclosed masses. 
New integrated light kinematics, unaffected by this bias, confirmed these results, showing a steady rise in the velocity dispersion from 360 parsecs into the centre.
Accounting for crowding in prior datasets gave velocity dispersions that matched the integrated light measurements.

An almost unambiguous signature of a black hole is a keplerian potential dominating over the potential of the galaxy.
Different assumptions for the shape of the dark matter halo and radius of tidal disruption for the galaxy were tested through orbit-based dynamical modelling, but all models consistently excluded the no black-hole hypothesis at over 95\% significance. 

Leo I represents an ideal environment in which to test our model. It is a dwarf spheroidal galaxy with a low gas content and a core stellar density at least 
two orders of magnitude less dense than that of a typical globular cluster. \cite{Ruiz_Lara_2020} find that the core of Leo I has a central density on the order of $0.7 \,\text{stars} \ \text{pc}^{-3}$, between 2-3 orders of magnitude less dense than the centres of typical globular clusters \citep{Gratton_2019}. In terms of definitions, the core of Leo I can be (marginally) described as an NSC - see for example Figure 2 from \cite{Stone_2017}. However, its central mass densities put Leo I at the very lowest end of the NSC spectrum and several orders of magnitude below that required for an NSC which can dynamically generate a MBH \citep[e.g.][]{Miller_2012, Stone_2017}.

%%%%%%%%%%%%%%%%%%%%%%%%%%Figure 1%%%%%%%%%%%%%%%%%%%%%%%%%%%%%%%%%%%%%%%%%%%%%
\begin{figure*}
\centering
\begin{minipage}{175mm}      \begin{center}
\centerline{
    \includegraphics[width=6cm, height=13cm, angle=-90]{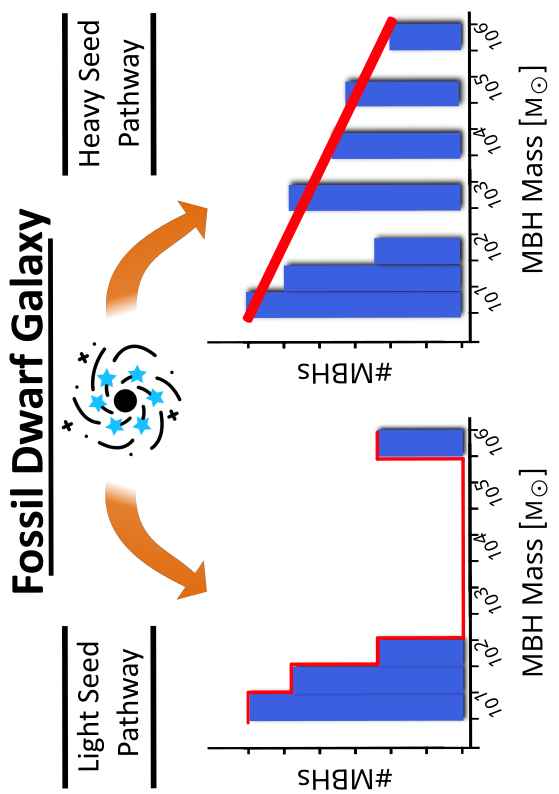}}
\caption{
The spectrum of black hole masses inside a fossil dwarf galaxy. For the \textit{light} seed pathway (left hand side) only one (central) MBH is expected with a large mass
gap between the mass of the central MBH and that of the stellar remnant black holes that will populate the galaxy. For the \textit{heavy} seed channel on the other-hand (right hand side)
a MBH continuum is expected as the gas initially fragments during the initial seeding process leaving behind a number of \textit{heavy} seed fragments. Some fragments will merge with the central objects — other fragments will remain in the core of the galaxy as passive MBHs. The detection of a continuum in black hole masses - particularly masses in the range $10^{3 - 5}$ \msolar - would represent very strong evidence of heavy seeding channel.} 
\label{Fig:Cartoon}
\end{center} \end{minipage}
\end{figure*}
%%%%%%%%%%%%%%%%%%%%%%%%%%%%%%%%%%%%%%%%%%%%%%%%%%%%%%%%%%%%%%%%%%%%%%%%%%%%%%%%%%

\section{Model} \label{Sec:Model}
\noindent Our model for determining the progenitor seeds of MBHs explores the seeding and growth of \textit{light} and \textit{heavy} seeds. 

\vspace{-0.2cm}
\subsection{Light Seed Growth \& Dynamics}
\noindent 
Both semi-analytic models and numerical simulations attempting to model the growth over cosmic time of PopIII remnant black holes (M$_{BH} \lesssim 10^3$ \msolarc) have consistently shown that these \textit{light} seeds do not grow \citep{Johnson_2007, Volonteri_2008, Alvarez_2009, Pacucci_2017_fb, Smith_2018}.
\textit{Light} seed growth has been shown to be possible within more idealised settings — particularly where it is able to accrete within the confines of a dense stellar cluster at high redshift \citep{Miller_2012}.
Pioneering work by \cite{PortegiesZwart_2004} demonstrated that stellar collisions in dense clusters can produce massive stars which in-turn collapse into MBHs - or perhaps also populating the pair instability mass gap with black holes \citep{Gonzalez_2021}. In a similar way \cite{Miller_2012}, \cite{Stone_2017} and \cite{Fragione_2022} have identified NSCs with velocity dispersions of greater than 40 \kms as ideal sites in which to grow black holes (via tidal captures and tidal disruptions) past an initial bottleneck and 
up to a point where gas accretion can take over. \\
\indent Others have investigated the growth of \textit{light} seeds, predominantly via gas accretion within dense environments \citep[e.g.][]{Alexander_2014, Lupi_2016, Natarajan_2021, Fragione_2022} as a possible pathway to growing initally
"light" black holes. \\
\indent However, of particular relevance to this letter, such a dense environment is not necessarily present in all dwarf galaxies and certainly not in the dwarf galaxy Leo I — the case study used in this paper. \\
\indent Nonetheless, we cannot exclude the possibility of \textit{light} seed rapid growth (through accretion) even in the environs of dwarf spheroidal galaxies like Leo I. 
We quantify the probability of a PopIII remnant black hole growing through accretion in the core of a galaxy as follows.
We first assume that the mass of the PopIII remnant is 500 \msolar (which is in itself an optimistic assumption), giving a Bondi-Hoyle radius (from which a cross-section can be calculated) of $R_{\rm{Bondi}} \sim 10^{-2}$ pc.
Firstly, the probability that a black hole finds itself in a sufficiently dense volume relative to the volume of the galactic core is
\begin{equation}
\rm{P_{BH\_in\_cloud} = (R_{cloud}/R_{galcore})^3} \, ,
\end{equation}
where $R_{\rm{cloud}}$ is the radius of the gas cloud and $R_{\rm{galcore}}$ is the radius of the core of galaxy. We set $R_{\rm{cloud}} = 0.1$ pc and 
$R_{\rm{galcore}} = 20$ pc. We then multiply this number by the number of clouds expected in this region. For this purpose, we assume that $1 \times 10^{-4}$ (1\% by volume) of  $R_{\rm{galcore}}$ is filled with sufficiently dense gas giving $N_{\rm{clouds}} \sim 800$. The values used here are based on the properties of the gas rich star forming galaxy found in \cite{Regan_2020b}.

Finally, we compute, assuming that the black hole walks a random trajectory around that galaxy, that the fraction of the volume sampled by the black hole, $V_{\rm{sampled}}$, in a Hubble time, 
$\tau_{\rm{Hubble}}$, is given by 
\begin{equation}
    \rm{V_{\rm{sampled}} = {{\tau_{Hubble} R_{Bondi}^2 v_{BH}} \over {2 R_{galcore}^3}}} \, ,
\end{equation}
where $\rm{v_{BH}}$ is the average relative velocity of the black hole (set here to be equal to the sound speed of the gas, $\sim 10$ \kms). The total probability of a single PopIII remnant accreting within a high-z galaxy (for which these numbers are derived) is then given by
\begin{equation} \label{probability}
    \rm{P_{growth} = P_{BH\_in\_cloud} \times N_{clouds} \times V_{sampled}} \, .
\end{equation}
 Using the canonical set of values noted above, which are consistent with gas rich early galaxies, Eqn \ref{probability} gives a probability that a 
 stellar mass black holes intersects a single dense gas cloud within a Hubble time as $P_{\rm{growth}} \sim 9 \times 10^{-8}$.

Given this estimate, the probability of two (or more) black holes within the same environment experiencing growth becomes infinitesimally small. This is just the probability of a single black hole encountering such a sufficiently dense environment once — when in reality a black hole must encounter such an environment on multiple (perhaps hundreds of) occasions. \\ 
\indent In short, unless \textit{light} seeds find themselves within a very dense environment in which growth becomes much more likely via dynamical processes, then \textit{light} seeds are extremely unlikely to grow. 
\vspace{-0.2cm}
\subsection{Heavy Seed Growth \& Dynamics}
\noindent Our assumptions on the mass of \textit{heavy} seeds are given by state-of-the-art cosmological simulations undertaken by numerous groups. The general agreement is that MBH seeds within the range $M_{\mathrm{seed}} = 10^3 - 10^5$ \msolar are possible \citep{Hosokawa_2013, Latif_2013d, Regan_2014a, Inayoshi_2014, Inayoshi_2014b, Latif_2016a, Regan_2018a, Regan_2018b}. In idealised settings, a single object (with masses up to $10^5$ \msolarc) can be formed \citep{Inayoshi_2014}, but for models in which more cosmologically consistent treatments are performed the formation and retention of multiple fragments is either moderate \citep[e.g.][]{Regan_2018a, Regan_2018b, Latif_2022} or more widespread \citep{Wise_2019, Regan_2020b}. While some of these fragments may eventually merge or be ejected from the halo, it is also likely that many will survive as isolated MBHs or in stable binaries. 

Current models for heavy seed formation suggest that several heavy seeds could form at the same time. Here, we show that if this is the case, then it is unlikely that all of them will merge with the central MBH. Hence, we propose that a signature of heavy seed formation in quiescent (i.e. those who have had no major mergers) dwarf galaxies is the detection of off-centered, wandering MBHs (see Figure \ref{Fig:Cartoon}) with masses in the range $M_{\rm{MBH}} = 10^3 - 10^5$. These MBH ``leftovers'' are the observational
signature of a \textit{heavy} black hole formation pathway in fossil dwarf galaxies. This signature does not apply to 
more massive galaxies in which MBHs can be incorporated through subsequent mergers over cosmic time -, nor does it (likely) apply to dwarf galaxies with high central densities typical of NSCs \citep{Stone_2017}. Although dynamical pathways (which straddle the definition of 
light and heavy seeds), may not create the continuum of MBHs we outline next, with instead a single MBH predicted to form within a dense system \citep[e.g.][]{Gonzalez_2021}. Instead the signature of an initial burst of heavy seeds will be a radial continuum of black hole masses as we now outline. \\
\indent We now explore through a simple analytic model how the impact of dynamical friction can lead to a fraction of the initial heavy seed population surviving within the fossil dwarf galaxy. Our goal is to demonstrate the existence of a MBH mass spectrum within a \textit{heavy} seed environment. We do not attempt a detailed exploration of the dynamics of MBH evolution as this is outside the scope of this letter (but see McCaffrey et al. in prep). \\
\indent To illustrate the existence of a MBH mass spectrum, we first calculate the dynamical friction \citep{Chandrasekhar_1943} timescale of a sample of \textit{heavy} seed masses born at different radii from the galactic centre. Using the formalism from \cite{Bar_2022} (which was originally applied to globular cluster sinking timescales in dwarf galaxies), we estimate the time for a MBH to sink to the centre of a dwarf galaxy as 
\begin{align} \label{Eqn:Tdf}
    \tau_{DF} &= \rm{{{v_{MBH}^3} \over {4 \pi G^2 \rho M_{MBH} C}}}\\
    &\approx \rm{2 \Big( \frac{\rm{v_{MBH}}}{10 \ \rm{km \ s^{-1}}} \Big) \Big(\frac{3 \times 10^6 \ M_{\odot} \ kpc^{-3}}{\rho}\Big) \Big(\frac{3 \times 10^5 \ M_{\odot}}{M_{MBH}}\Big) \ \frac{2}{C}} \ \rm{Gyr} \, ,
\end{align}
where $\rho$ is the background density of the medium inducing the dynamical friction, $M\rm{_{MBH}}$ is the mass of the MBH, $\rm{v_{MBH}}$ is the relative velocity of the MBH, and $C$ is a dimensionless factor accounting for the velocity dispersion of the medium and the Coulomb logarithm \citep{Hui_2017}.

\begin{figure}
\centering
\centerline{
    \includegraphics[width=8.6cm, height=5cm]{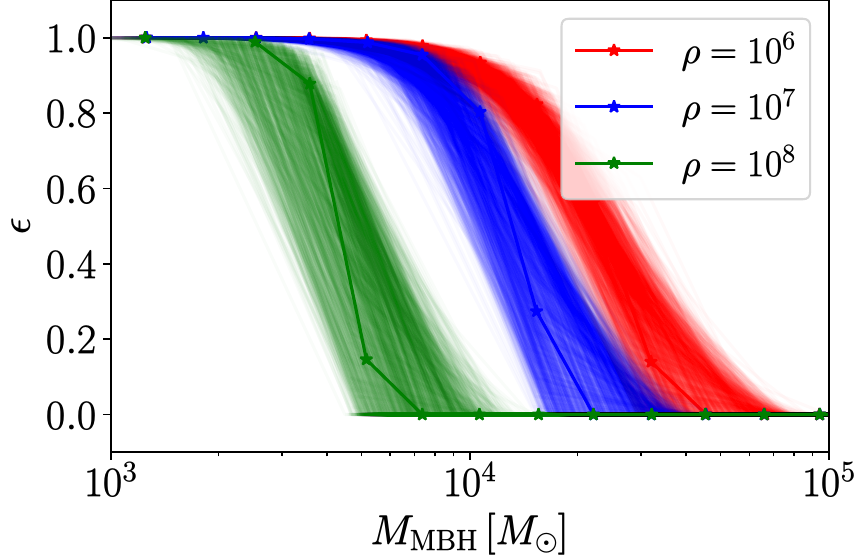}}
\caption{The survivor fraction, $\epsilon$, as a function of the black hole mass, M$_{MBH}$. The background density is varied from $\rho = 10^6$ \msolarc/kpc$^3$ up to 
$\rho = 10^8$ \msolarc/kpc$^3$ (we skip the units in the legend). Above $\rho = 10^8$ \msolarc/kpc$^3$ the density starts to become close to that found in globular clusters and hence much denser than a typical dwarf spheroidal galaxy like Leo I. For the lower average density range (i.e. $\rho \sim 10^6$ \msolarc/kpc$^3$) the survival rate of MBHs with M$_{MBH}\lesssim10^5$ \msolar is non-zero. As the background density increases, the dynamical friction force becomes stronger gradually pulling all masses towards the centre.}
\label{Fig:Epsilon}
\end{figure}

Using this value for the dynamical friction time, $\tau_{DF}$, the radius, $R$, to which the MBH sinks after a time $t$, (assuming a core halo profile) can be estimated from \cite{Bar_2021} using:

\begin{equation} \label{Eqn:NewRadius}
    R = r_0 \ \rm{exp} \Big (-\frac{t}{2 \tau_{DF}} \Big) \, ,
\end{equation}
where we set $r_0 = 200$ pc (as an approximate virial radius for a canonical dwarf galaxy) and $t = \tau_{Hubble}$.
Finally, using the value of the new radius, $R$, we can now estimate the MBH merger rate, $\Gamma$, as \citep{Bar_2022}
 
\begin{equation} \label{Eqn:MergerRate}
      \Gamma (R) \simeq \rm{\kappa_{MBH} \ \sigma_{MBH} \ v_{MBH}} \, ,
\end{equation}
where $\kappa_{MBH}$ is the number density of initial \textit{heavy} seeds calculated at the new radius $R$, $\sigma_{MBH}$ is the cross section for becoming gravitationally bound\footnote{we note here that the approximation of being gravitationally bound does not imply that the black holes will necessarily merge\citep{Begelman_1980, Lodato_2009} but is nonetheless a conservative approximation} ($\sigma_{MBH} = \pi R_{Bondi}^2$) and $\rm{v_{MBH}}$ is the relative velocity of the MBH (which we set equal to the sound speed). To calculate $\rm{\kappa_{MBH}}$, we divide the number of initial \textit{heavy} seeds, $\rm{N_I}$ by the volume (i.e. $4/3 \, \pi R^3$). We set $\rm{N_I} = 20$ based on the results of \cite{Regan_2020b}.
We are interested in the survivor fraction, $\epsilon$, not in the number of mergers, $\rm{N_{MBH} = \Gamma (R) \times N_I \times \tau_{Hubble}}$. Specifically, $\epsilon$, defined between 0 and 1, is the fraction of MBHs that survive and 
do not merge with another MBH and instead orbit the galactic centre at some radius $R$. $\epsilon$ is given by 

\begin{equation}
    \epsilon = \rm{1 - {{N_{MBH}} \over {N_I}}} \, 
\end{equation}
To illustrate this model we run Monte-Carlo simulations of the above scenario and plot the results in Figure \ref{Fig:Epsilon}. \\
\indent For our Monte-Carlo model we sample from a normal distribution of \textit{heavy} seed masses with a mean of $1.5 \times 10^4$ \msolar and a standard deviation of 0.45. The distribution of heavy seeds is unknown (assuming they exist in the first place) and this distribution is chosen based on  \cite{Regan_2020b}. Our results are not sensitive to the details of the distribution but do rely on initial fragmentation and the production of multiple 
heavy seeds within the parent gas cloud. 
We modify the background density parameter, $\rho$, to illustrate how the survivor fraction, $\epsilon$, can vary as a function of MBH mass and background
density. An accurate calculation of the sinking timescale is non-trivial and depends on detailed knowledge of the dwarf galaxy environment, including the cusp/core density profile and the time evolution of the galaxy \citep[e.g.][]{Weinberg_2015, Sanchez_2006, Sanchez_2022, Shao_2021}. As a result, we parameterise these unknown variables by varying the background density. The stellar density in LeoI is 
approximately $10^7$ \msolarc/kpc$^3$ with other dwarf galaxies in the local group having values varying around this figure by several dex \citep{McConnachie_2012}. High-z dwarf galaxies tend to be more gas rich and can have densities at the higher end of our parameterisation but
their centres are also highly dynamic and simulations have consistently shown that MBHs struggle to sink towards the galactic "centre" 
\citep[e.g.][]{Pfister_2017, Lescaudron_2022}.\\
\indent Figure \ref{Fig:Epsilon} shows the impact of different background densities on the survival fraction of MBHs. For background densities of $\rho \gtrsim 10^8$ \msolar kpc$^{-3}$ (green line, similar to the density inside globular clusters), the survivor fraction drops rapidly above the \textit{heavy} seed threshold (M$\rm{_{MBH}} \gtrsim 10^3$ \msolarc) i.e., most heavy seeds merge through mass segregation. However, for values of the background density parameter closer to that expected in typical dwarf spheroidal galaxies, the survivor fraction remains high ($\epsilon \gtrsim 0.5$) up to relatively high MBH masses (M$\rm{_{MBH} > 10^4}$ \msolarc). For background density parameters of $\rho \sim 10^6$ \msolar kpc$^{-3}$ (red line, similar to the typical background density found outside the core of Leo I), the survivor fraction is non-zero out to M$_{MBH} \gtrsim 6 \times 10^4$ \msolarc. Our model cannot account for the growth experienced by black holes over time and 
hence we are therefore assuming that these seeds do not grow. This is likely to be
a very good assumption for all black holes with masses M$_{MBH} \lesssim 10^5$ \msolar as numerical simulations with realistic seeding prescription show that black holes below this mass scale show little or no growth \citep[e.g.][]{DiMatteo_2022}. Above this mass scale black holes may sink and grow more efficiently.\\
\indent This admittedly simplified calculation shows that for density parameterisations typical of dwarf spheroidal galaxies, there is a large window in the \textit{heavy} seed mass spectrum 
($10^3 \ \rm{M_{\odot} \lesssim M_{MBH} \lesssim 10^5 \ M_{\odot}}$) for which the survivor fraction is non-zero. The most massive \textit{heavy} seeds can readily sink to the centre — the estimated dynamical
mass of the MBH at the centre of Leo I is M$_{MBH} \sim 3 \times 10^6$ \msolar (see \S \ref{Sec:LeoI}). Left behind on off-nuclear orbits are \textit{heavy} seeds likely formed during the same formation epoch as the most massive seed but which have not yet sunk to the centre, due to their lower masses. In contrast to the wandering MBH paradigm typically discussed in the literature \citep[e.g.][]{Tremmel_2018, Reines_2020, Mezcua_2020, Bellovary_2021, Greene_2021, Weller_2022} these MBHs form in-situ (i.e. not acquired via mergers) and slowly sink toward the centre of the galaxy. Their intrinsically different masses result in a diversity of timescales to sink and merge, hence their very existence results in a unique signature of their formation pathway. Additionally, we may in practice be somewhat conservative in our analysis here since we are assuming pure `Chandrasekhar' style dynamical friction. However, 
it is well known that the inspiral time may in fact be much longer \citep{Read_2006, Goerdt_2006}. In that case our results be a lower limit and the true survival fraction, $\epsilon$, is likely to be higher. As a final note on the distribution of these survivors it may be, depending on the composition of the core, that the 
mass distribution becomes inverted to what might be naively expected. \cite{Kaur_2018} has shown that core stalling can lead to positive mass dependence of radial sinking versus mass such that
$R_s \sim M_{MBH}^{1/5}$ where $R_S$ is the filtering radius. In this case the more massive black holes may reside further from the centre \citep{Kaur_2022}.
%%%%%%%%%%%%%%%%%%%%%%%%%%Figure 2%%%%%%%%%%%%%%%%%%%%%%%%%%%%%%%%%%%%%%%%%%%%%
%%%%%%%%%%%%%%%%%%%%%%%%%%%%%%%%%%%%%%%%%%%%%%%%%%%%%%%%%%%%%%%%%%%%%%%%%%%%%%%%%%
\section{Discussion, Conclusions \& Future Observational Markers} \label{Sec:Discussion}
\noindent The discovery \citep{Bustamante-Rosell_2021} of a MBH at the centre of the dwarf spheroidal galaxy Leo I is remarkable in many ways. Leo I has an estimated virial mass of M$_{vir} = (7 \pm 1) \times 10^8$ \msolar \citep{McConnachie_2012} and a stellar mass of M$_{*} = 5.5 \times 10^6$ \msolar \citep{Mateo_2008}.
With M$_{MBH} = (3.3 \pm 2) \times 10^6$ \msolarc, this black hole is significantly over massive, by a factor $\sim 10^3$, compared to the virial mass of the halo. What are the consequences for the formation pathways of the central MBH?\\
\indent Numerous authors have argued that satellite galaxies irradiated by a nearby massive galaxy will host over massive 
MBHs \citep{Agarwal_2013, Natarajan_2017, Scoggins_2022} formed via the \textit{heavy} seed paradigm in which
super-massive stars are one potential intermediate stage. For the case of Leo I, the \textit{heavy} seed formation pathway
may have been induced via either an intense burst of Lyman-Werner radiation, the rapid assembly of 
the original Leo I galaxy component, baryonic streaming velocities — or a combination of 
one or more of these mechanisms. In either case, the result is broadly similar: a small number of MBHs are expected to form in the centre of the embryonic dwarf galaxy, Leo I in this case, with some surviving to the present epoch. \\
\indent Dwarf galaxies are potential sites to search for the fossils 
of the very early stages of MBH formation \citep{Volonteri_2008, VanWassenhove_2010}. Here, we 
extend that idea by also suggesting that a specific observational signature of \textit{heavy} seed
MBH formation would be the existence of a continuum in mass of MBHs, from stellar mass to the mass of the central MBH. The continuum being made up of the population of stellar mass black holes formed from the end point of stellar evolution plus an additional, smaller, component made up from an initial burst of heavy seed formation.
Figure \ref{Fig:Cartoon} illustrates this paradigm and its outcome. If the seed for the central MBH was a light seed, then no such continuum should exist, and there should be a clear gap in the black hole mass spectrum in fossil dwarf galaxies between the mass of the most MBH in the galaxy and the 
population of stellar mass black holes. In this case a single 
light seed grows spectacularly through accretion but the process is 
sufficiently rare that only a single object emerges from the population of light seeds. \\
\indent While the black holes carry no information of their accretion or merger history that is easily disentangled \citep{Pacucci_2020a}, there may be clues 
from the black hole demographics inside fossil dwarf galaxies like Leo I. Fragmentation, even in the \textit{heavy} seed formation channel, is a 
robust prediction. As we demonstrate in \S \ref{Sec:Model}, at least some of the original MBHs will survive as isolated or binary MBHs. It is these leftover MBHs, with masses less than that of the central MBH, that we highlight as observational signatures of a \textit{heavy} seed formation scenario. \\
\indent It is essential to note that the 
absence of a continuum of black hole masses does not by itself falsify the \textit{heavy} seed scenario, as mergers, ejections, or very low levels of fragmentation could equally be responsible. Instead, detecting a black hole mass spectrum would be strong evidence for a \textit{heavy} seed formation channel. 

A final unknown remains: what are the signatures of off-nuclear MBH in dwarf galaxies, and — most importantly — are they detectable at all? Electromagnetic emission from accretion onto MBHs in relic dwarfs such as Leo I is expected to be faint, because of the lack of gas. MBHs wandering outside the central regions of galaxies are now routinely discovered, also in dwarf galaxies (see, e.g., \citealt{Reines_2020, Greene_2020, Greene_2021}), with simulations showing that the presence of off-centered MBHs should be the norm in dwarfs (due to the long inspiral times) \citep{Bellovary_2021}. Recently, \cite{Seepaul_2022} showed that wandering MBHs in the Milky Way galaxy, or in close-by galaxies as Leo I, should be detectable in a wide range of frequencies, pending the presence of a minimum density of gas to trigger advection-dominated accretion flows \citep{Pacucci_2022}. Alternatively, the merger of MBHs could be studied by third-generation (3G) gravitational wave observatories, such as the Einstein Telescope \citep{Maggiore_2020} and the Cosmic Explorer \citep{Reitze_2019}, with the added advantage of a wide redshift range, crucial in building up the statistics necessary to probe demographics. In fact, \cite{Valiante_2021} and \cite{Chen_2022} recently investigated the merger of MBHs in the mass range of our interest.

We encourage further in-depth observations 
and modelling of the dynamics inside Leo I and similar dwarf galaxies as an ideal environment in which to probe MBH seeding channels. 

\vspace{-0.5cm}
%====================================================================
\section*{Acknowledgements}
%====================================================================
\noindent The authors wish to thank Nick Stone, Giacomo Fragione and  Vivienne Baldassare for reading and providing feedback on early drafts of the text. J.R. acknowledges support from the Royal Society and Science Foundation Ireland under grant number URF$\backslash$R1$\backslash$191132. JR also thanks the organisers of the \textsc{Intermediate Mass Black Holes: New Science from Stellar Evolution to Cosmology} during which the concept for this paper was born.
F.P. acknowledges support from a Clay Fellowship administered by the Smithsonian Astrophysical Observatory. This work was also supported by the Black Hole Initiative at Harvard University, funded by grants from the John Templeton Foundation and the Gordon and Betty Moore Foundation. We thank the anonymous referee for a constructive report.
%%%%%%%%%%%%%%%%%%%%%%%%%%%%%%%%%%%%%%%%%%%%%%%%%%
\section*{Data Availability}
Data generated in this research will be shared on reasonable request to the corresponding author.

%%%%%%%%%%%%%%%%%%%% REFERENCES %%%%%%%%%%%%%%%%%%

\bibliographystyle{mnras}
\bibliography{Leo1}

\bsp	
\label{lastpage}
\end{document}